\documentclass[article]{revtex4}

    \usepackage{amsmath}
    \usepackage{makeidx}
    \usepackage{amsfonts}
    \usepackage[ansinew]{inputenc}
    \usepackage[usenames,dvipsnames]{pstricks}
    \usepackage{subfigure}
    \usepackage{epsfig}
    \usepackage{pst-grad} 
    \usepackage{pst-plot} 
    \usepackage[colorlinks,hyperindex]{hyperref}
    \hypersetup
    {
        colorlinks,%
        citecolor=black,%
        linkcolor=black,%
        urlcolor=black,%
    }

\newcommand{\be}{\begin{equation}}
\newcommand{\ee}{\end{equation}}
\newcommand{\bea}{\begin{eqnarray}}
\newcommand{\eea}{\end{eqnarray}}

\preprint{YITP-13-113}

\makeindex

\begin{document}

\title{Non perturbative effects of primordial curvature perturbations on the apparent value
of a cosmological constant}
\author{Antonio Enea Romano$^{1,2,5}$, Sergio Sanes$^{5}$, Misao Sasaki$^{1}$, Alexei A. Starobinsky$^{3,4}$}
\affiliation{
${}^{1}$Yukawa Institute for Theoretical Physics, Kyoto University,
Kyoto 606-8502, Japan;\\
${}^{2}$Department of Physics, McGill University, Montr\'eal, QC H3A 2T8,
Canada; \\
${}^{5}$Instituto de Fisica, Universidad de Antioquia, A.A.1226, Medellin,
${}^{3}$Landau Institute for Theoretical Physics RAS, Moscow 119334, Russia; \\
${}^{4}$Research Center for the Early Universe, Graduate School of
Science, The University of Tokyo, Tokyo 113-0033, Japan }

\begin{abstract}

We study effects on the luminosity distance of a local
inhomogeneity seeded by primordial curvature perturbations of the
type predicted by the inflationary scenario and constrained by the
cosmic microwave background radiation.
We find that a local underdensity originated from a
one, two or three standard deviations peaks of the primordial
curvature perturbations field can induce corrections to the value
of a cosmological constant of the order of $0.6\%,1\%,1.5\%$
respectively. These effects cannot be neglected in the precision
cosmology era in which we are entering. Our results can be considered an upper bound for the effect of the monopole component of the local non linear structure which can arise from primordial curvature perturbations and requires a fully non perturbative relativistic treatment.
\end{abstract}

\keywords{Cosmological constant, Inflation, Cosmology}
\pacs{98.80.Es, 98.65.Dx, 98.80.-k}

\maketitle
\section{Introduction}
Before the invention of the inflationary scenario of the early
Universe, the standard cosmological model was based on the
assumption of global large scale homogeneity and isotropy (often
dubbed "the Cosmological principle") which reflects itself in the
observed approximate isotropy of the Hubble flow and the
temperature of the cosmic microwave background (CMB) radiation.
With the inflationary scenario, there is no more need in such
assumption since viable realizations of this scenario {\em
predict} the large-scale space-time metric to be approximately
homogeneous and isotropic up to very large (but still finite)
scales fantastically exceeding the present Hubble radius. In
addition, these viable inflationary models predict a specific
structure of small inhomogeneous metric perturbations of the
scalar and tensor types inside the observed part of Universe
(inside the last scattering surface if speaking about photons).
These perturbations are supposed to be the seeds of the CMB
angular temperature anisotropy and polarization. Also, from the
scalar perturbations, galaxies and their clusters, as well as
other compact objects and the large-scale structure of  the
spatial distribution of galaxies in the Universe, have been
formed at later time due to gravitational instability. Due to the
quantum (in fact, quantum-gravitational) origin of these small
primordial perturbations in the inflationary scenario, they can be
very well described as classical stochastic quantities with the
almost Gaussian statistics. Because of this stochasticity, the
Cosmological principle is not exact even at observable scales less
than the Hubble radius, some deviations from it are certainly
expected, though of a specific type and sufficiently small
probability, e.g. near large extrema of perturbations. However,
cosmic variance  \cite{Marra:2013rba,Li:2008yj,Shi:1997aa,Wang:1997tp,2013A&A,Ben-Dayan:2014swa}  makes such a possibility viable and worth investigation.
It should be emphasized that the effects we study in this letter correspond to highly non linear structures which cannot be studied using a perturbative approach, since the relativistic correction can be dominant \cite{Bolejko:2012uj}, and for this reason they can differ from previous estimations of the effects of cosmic variance which were based on perturbative relativistic or Newtonian approximations.

That is why we study the effects of late time local inhomogeneities corresponding to large peaks of the primordial scalar (curvature) perturbations on the apparent values of parameters of the standard cosmological model, mainly on the value of a cosmological constant. We model a present day local inhomogeneity around such a large peak by the
Lemaitre-Tolman-Bondi (LTB) solution with the initial condition
compatible with inflation, i.e. with the homogeneous initial
curvature singularity. The spherical symmetry of the metric is
justified by the known property of large peaks of a Gaussian
random field \cite{Bardeen:1985tr} to be approximately spherically
symmetric, and we consider the case in which an observer is
located at the center of the peak.
In this way we are able to relate the local inhomogeneity to the
primordial curvature perturbation spectrum constrained by CMB
anisotropy observations. Previous studies have shown how ignoring
the presence of such a local inhomogeneity could lead to the wrong
conclusion of the presence of evolving dark energy
\cite{Romano:2010nc}, while in fact dark energy is simply a
cosmological constant. Similar approach was used in
\cite{Valkenburg:2011ty} and \cite{Marra:2012pj}. Here we consider the effects on
the supernovae Ia luminosity distance, and quantify the effect on
the estimation of the apparent value of the cosmological constant.
Other attempts to study the effects of inhomogeneities on
cosmological observables consisted in implementing some averaging
procedure \cite{Romano:2006yc, Romano:2009xw, Fanizza:2013doa,
BenDayan:2013gc, Buchert:2001sa, Buchert:2002ht} or to consider
inhomogeneities as alternative to dark energy
\cite{GarciaBellido:2008yq, February:2009pv, Uzan:2008qp, Clarkson:2007bc,
Zuntz:2011yb,Ishibashi:2005sj, Bolejko:2011ys, Romano:2009xw, Romano:2007zz,
Romano:2012yq, Romano:2009qx, Zibin:2011ma, Bull:2011wi,
Balcerzak:2012bv}.

Following the definition of apparent value of the cosmological
constant given in \cite{Romano:2011mx} we find that the effects of
a local overdensity are not very important, while a local
underdensity can lead to a correction of the apparent value of the
cosmological constant up to a order of $1.5\%$, which cannot be
ignored in high precision cosmology. While our approach here is
based on making a theoretical connection between the local
Universe today and early Universe physics, there have been some
recent direct observational evidences \cite{Keenan:2013mfa} which
we may actually live inside a local inhomogeneity, which could
have arisen from a primordial curvature perturbation peak of the
type we study here, as predicted by \cite{Bardeen:1985tr}. Other
possible evidences of  being located inside a local
inhomogeneity come from the apparent tension
between the estimation of cosmological parameters from local
observations and the Planck satellite results \cite{Verde:2013wza,
Ade:2013zuv}.
\section{Primordial curvature perturbations and late time
inhomogeneities}
The metric after inflation on scales much larger
than the Hubble scale can be written as
\be
ds^2=-dt^2 + a_F^2(t)e^{2\zeta({\bf r})}(dr^2+r^2d\Omega^2)\, \label{minf} \ee
where $r = |{\bf r}|$, and $\zeta({\bf r})$ may be interpreted as
a local, space-dependent number of $e$-folds $N$ from a
hypersurface of uniform spatial scalar curvature (called the flat
hypersurface) during inflation (up to a constant absorbable into
$a_F$). This relation is the basis of the so-called $\delta N$
formalism, first used in \cite{Starobinsky:1982ee} in the case of
a single field inflation, and then extended to multiple field
inflation in \cite{Starobinsky:1986fxa, Sasaki:1995aw}. Here we
neglect tensor perturbations (primordial gravitational waves)
since, first, their power is suppressed by at least one order of
the small inflationary slow-roll parameter $\varepsilon$ compared
to scalar (curvature) perturbations and, second, they are not
subjected to gravitational instability and their amplitude
decreases $\propto a_F^{-1}$ inside the Hubble radius. Near large
peaks of primordial perturbations, we approximate $\zeta =
\zeta(r)$. Such points certainly exist somewhere in space.

The Lemaitre-Tolman-Bondi (LTB) solution is a pressureless
spherically symmetric solution of Einstein's field equations given
by \cite{Lemaitre:1933qe, Tolman:1934za, Bondi:1947av}
\begin{eqnarray}
\label{LTBmetric} ds^2 = -dt^2  + \frac{\left(R,_{r}\right)^2
dr^2}{1 + 2\,E(r)}+R^2 d\Omega^2 \, ,
\end{eqnarray}
where $R$ is a function of the time coordinate $t$ and the radial
coordinate $r$, $E(r)$ is an arbitrary function of $r$, and
$R_{,r}=\partial_rR(t,r)$. The Einstein's equations with a
cosmological constant give
\begin{eqnarray}
\label{eq2} \left({\frac{\dot{R}}{R}}\right)^2&=&\frac{2
E(r)}{R^2}+\frac{2M(r)}{R^3}+\frac{\Lambda}{3} \, , \\
\label{eq3} \rho(t,r)&=&\frac{2M,_{r}}{R^2 R,_{r}} \, ,
\end{eqnarray}
where $M(r)$ is an arbitrary function of $r$, $\dot
R=\partial_tR(t,r)$ and $c=8\pi G=1$ and is assumed in the rest of
the paper. We will also adopt, without loss of generality, the
coordinate system in which $M(r)\propto r^3$, and fix the geometry
of the solution by using a function $k(r)$ according to
$2E(r)=-k(r)r^2$.

As shown in \cite{Romano:2010nc} it is possible to choose an
appropriate time when we can match the LTB metric and the metric
after inflation given in eq.(\ref{minf}). The result is a relation
between the primordial curvature perturbation and the function
$k(r)$ :
\begin{equation}
 k(r)=-\frac{1}{ r^2}[(1+r\zeta')^2-1] \,, \label{kz}
\end{equation}
which in the linear regime $r\zeta'\ll1$  reduces to \be
k(r)=-2\frac{\zeta'(r)}{r}\,. \ee The approximation of the
spherical symmetry is justified by the known property of a
Gaussian random field \cite{Bardeen:1985tr} according to which
large peaks of a stochastic function tend to have a spherical
shape. In the rest of the paper we will make the additional
assumption to be located at the center of such a spherically
symmetric inhomogeneity. This is supported by other evidences of
isotropy such as the cosmic microwave background (CMB) radiation,
implying that if any local inhomogeneity around us is actually
present, this should be highly spherically symmetric. In this
sense the effects we will consider are associated to the monopole
component of the local structure surrounding us, assuming this was
seeded by a  few $\sigma$ peak of the metric perturbation $\zeta$.
Such perturbation still corresponds to a small additional density
contrast if its size is sufficiently large.
The assumption of being located at the center of such a peak
allows us to set an upper bound on the magnitude of the effects on
the estimation of cosmological parameters, in particular on the
value of the cosmological constant.
\begin{figure}
\centering
\begin{tabular}{c}
\vspace{0in}
 \includegraphics[scale=0.13,angle=-90]{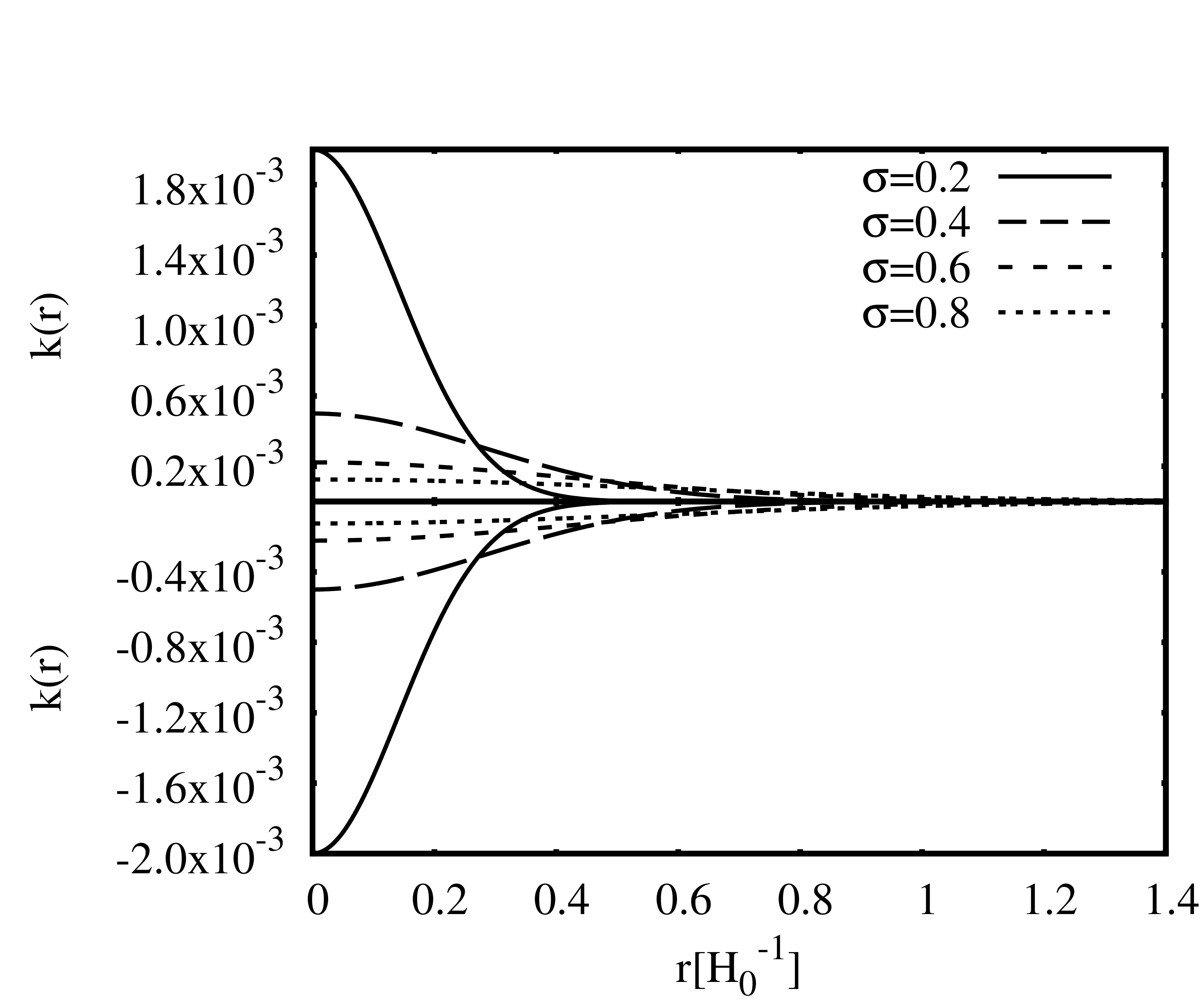}\\
 \vspace{0in}
 \includegraphics[scale=0.13,angle=-90]{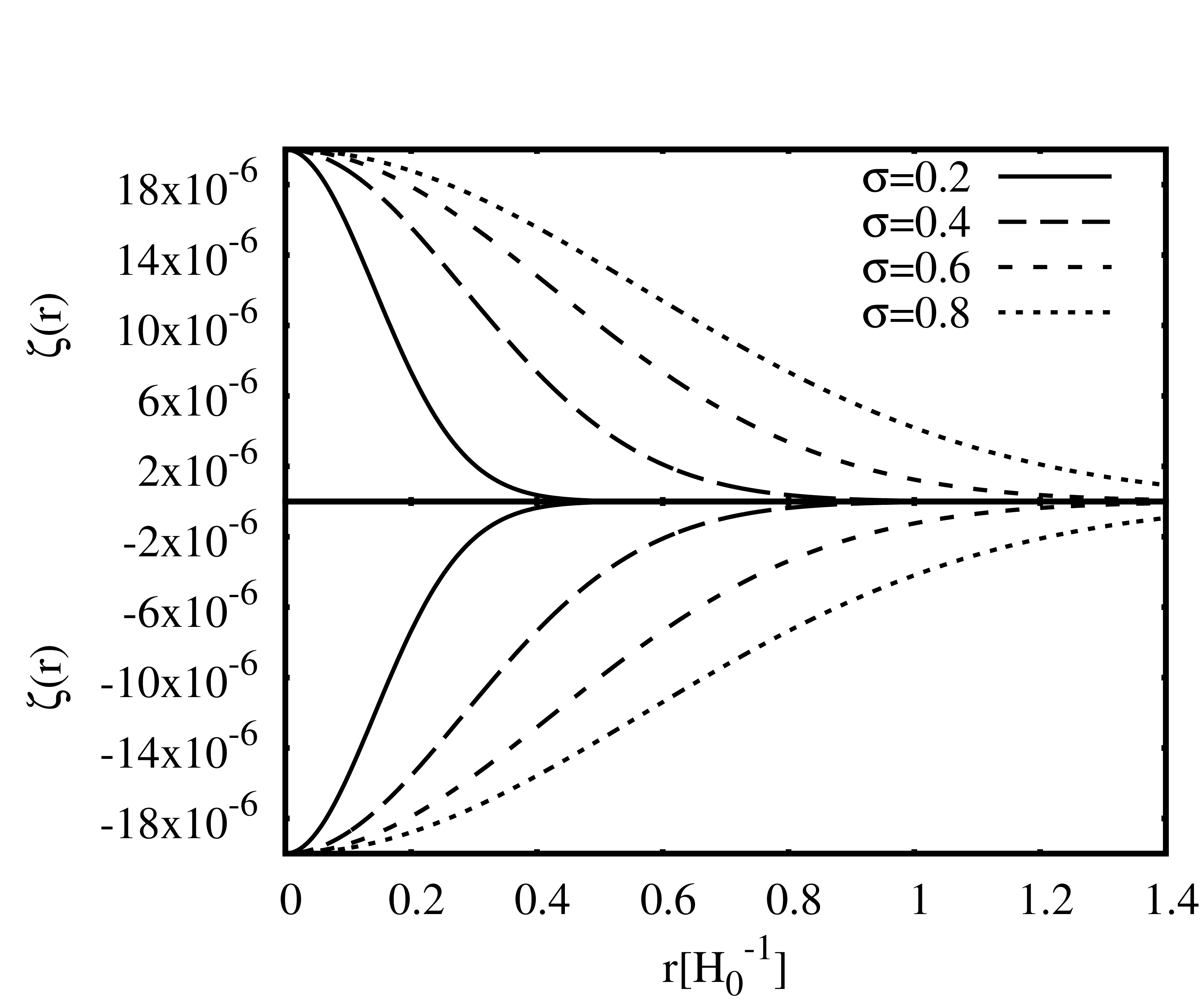}
\end{tabular}
\caption{In the top $k(r)$ is plotted for $A=2\times(5\times
10^{-5})$ (top) and $A=-2\times(5\times   10^{-5})$ (bottom), and
$\zeta(r)$  in the bottom for $A=2\times(5\times   10^{-5})$ (top)
and $A=-2\times(5\times   10^{-5})$ (bottom). Different lines
correspond to different values of $\sigma$,  expressed in units of
$H_0^{-1}$. } \label{fig-k-z}
\end{figure}

\begin{figure}
\centering
\begin{tabular}{c}
\includegraphics[width= 0.5 \columnwidth]
{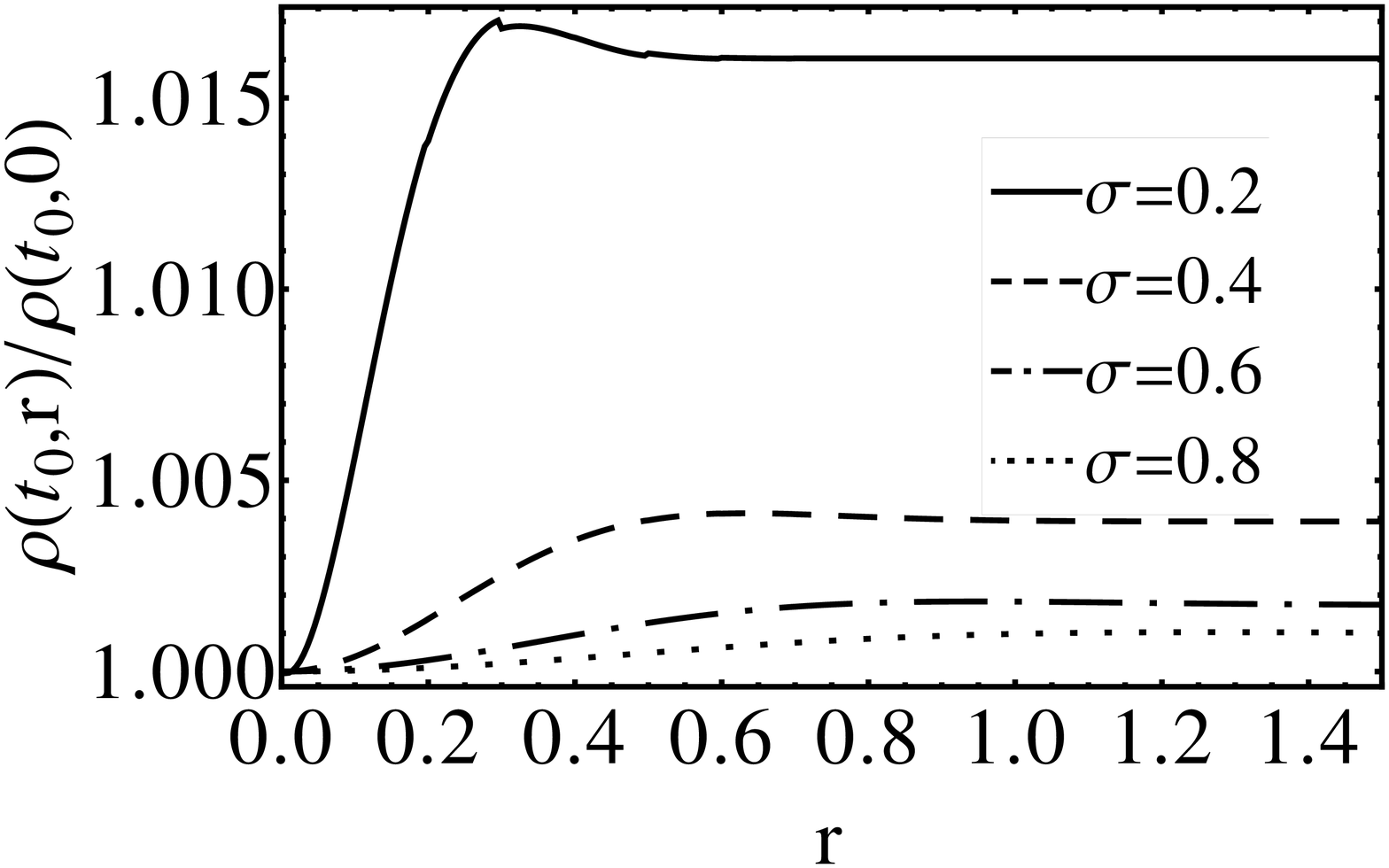}
\includegraphics[width= 0.5 \columnwidth]
{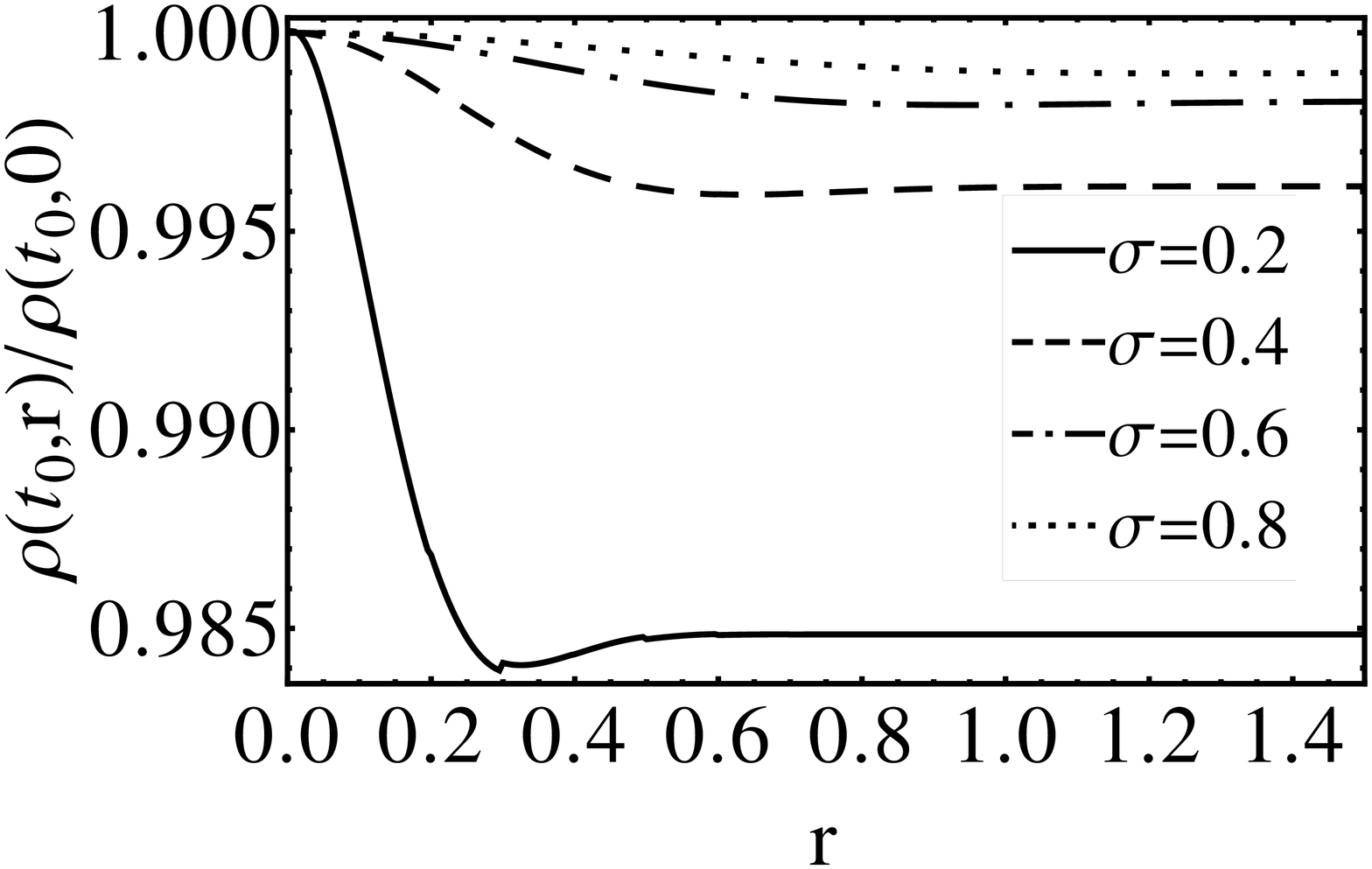}
\end{tabular}
\caption{The energy density ratio $\rho(t_0,r)/\rho(t_0,0)$ at the
time observation $t_0$  is plotted as function of the radial
coordinate for $A=-2\times(5\times   10^{-5})$ on the left and
$A=2\times(5\times   10^{-5})$ on the right. As can be seen
positive primordial curvature perturbations, correspond to a
central overdensity, and negative primordial curvature
perturbations correspond to a central underdensity. Another
important feature is that larger values of $\sigma$ correspond to
smaller levels of inhomogeneity. The radial coordinate $r$ and
$\sigma$ are expressed in units of $H_0^{-1}$.   } \label{rho}
\end{figure}

\begin{figure}
\centering
\begin{tabular}{c c}
 \includegraphics[width= 0.5 \columnwidth]
 {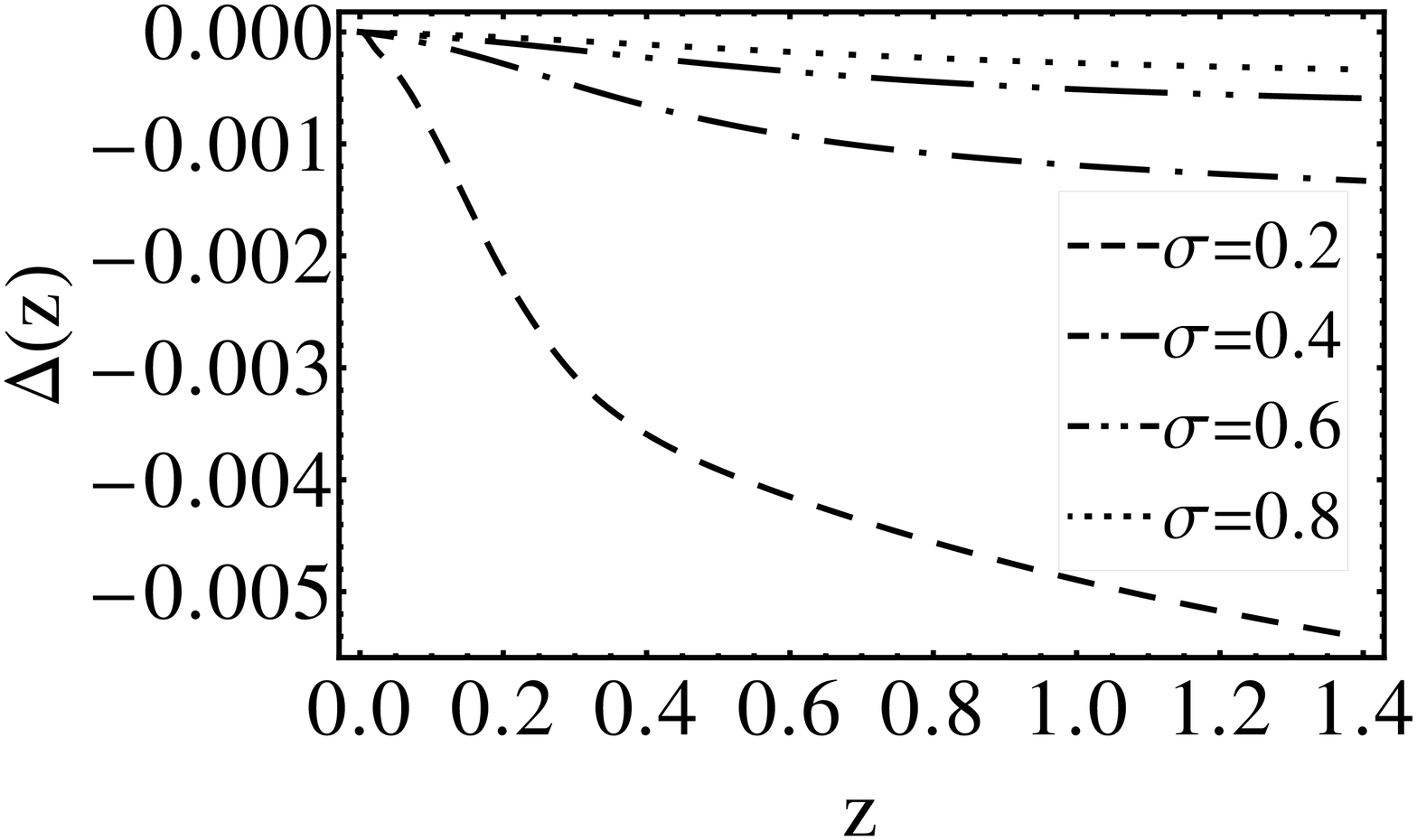}
 \includegraphics[width= 0.5 \columnwidth]
 {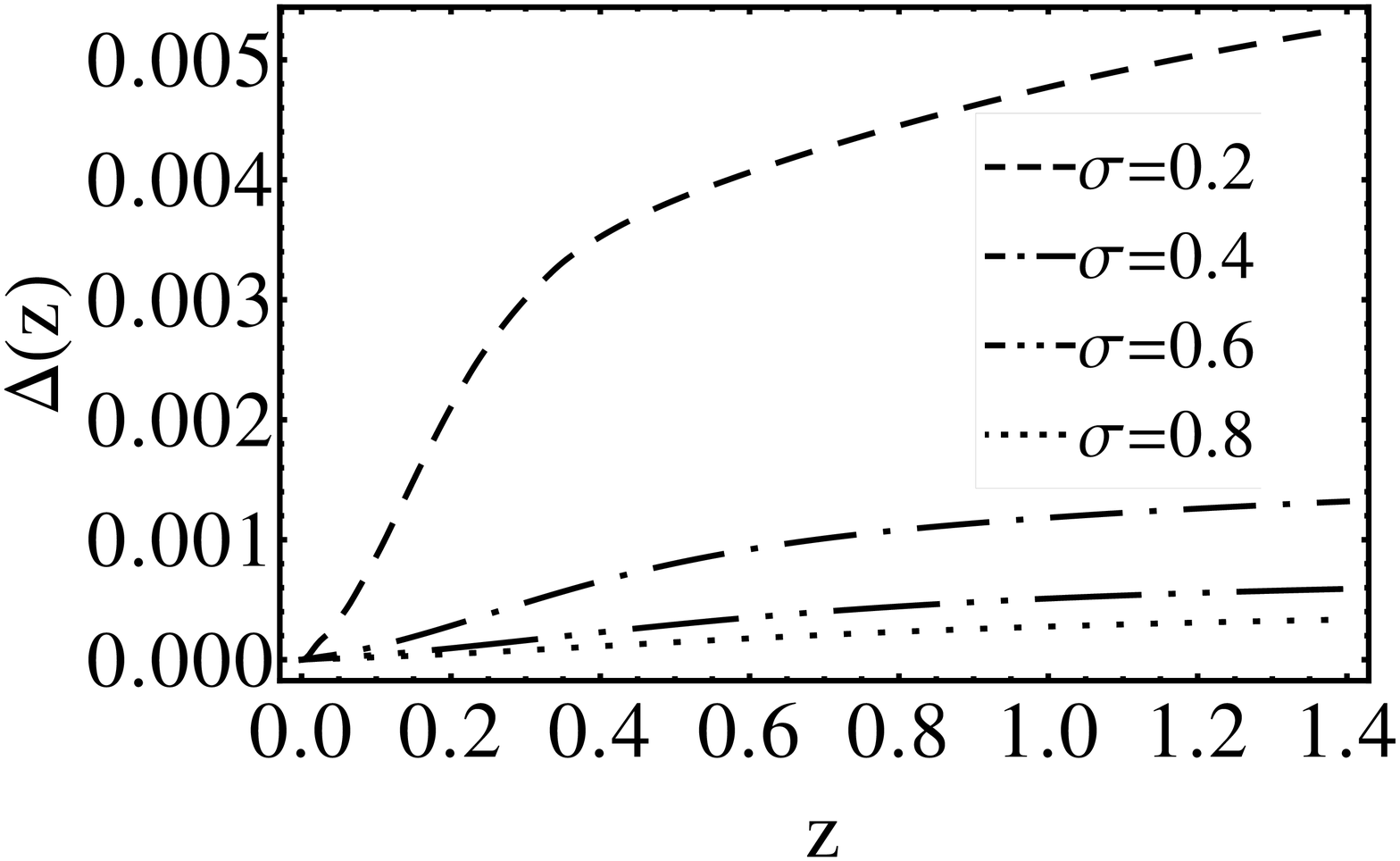}
\end{tabular}
\caption{The relative difference $\Delta(z)=(D_L^{\Lambda
CDM}(z)-D_L^{\Lambda LTB}(z))/D_L^{\Lambda CDM}(z)$ of the
luminosity distance between the $\Lambda LTB$ case and $\Lambda
CDM$ is plotted for different values of $\sigma$, where the latter
is in units of $H_0^{-1}$. The left figure corresponds to
$A=-2\times(5\times   10^{-5})$  and the right to
$A=2\times(5\times   10^{-5})$. A local underdensity,
corresponding to $A<0$, is associated to a larger luminosity
distance respect to the homogeneous case, while local
overdensities give a smaller distance. $\sigma$ is expressed in
units of $H_0^{-1}$.} \label{Dl}
\end{figure}

\begin{figure}
\centering
\begin{tabular}{c}
\includegraphics[width= 0.5 \columnwidth]
{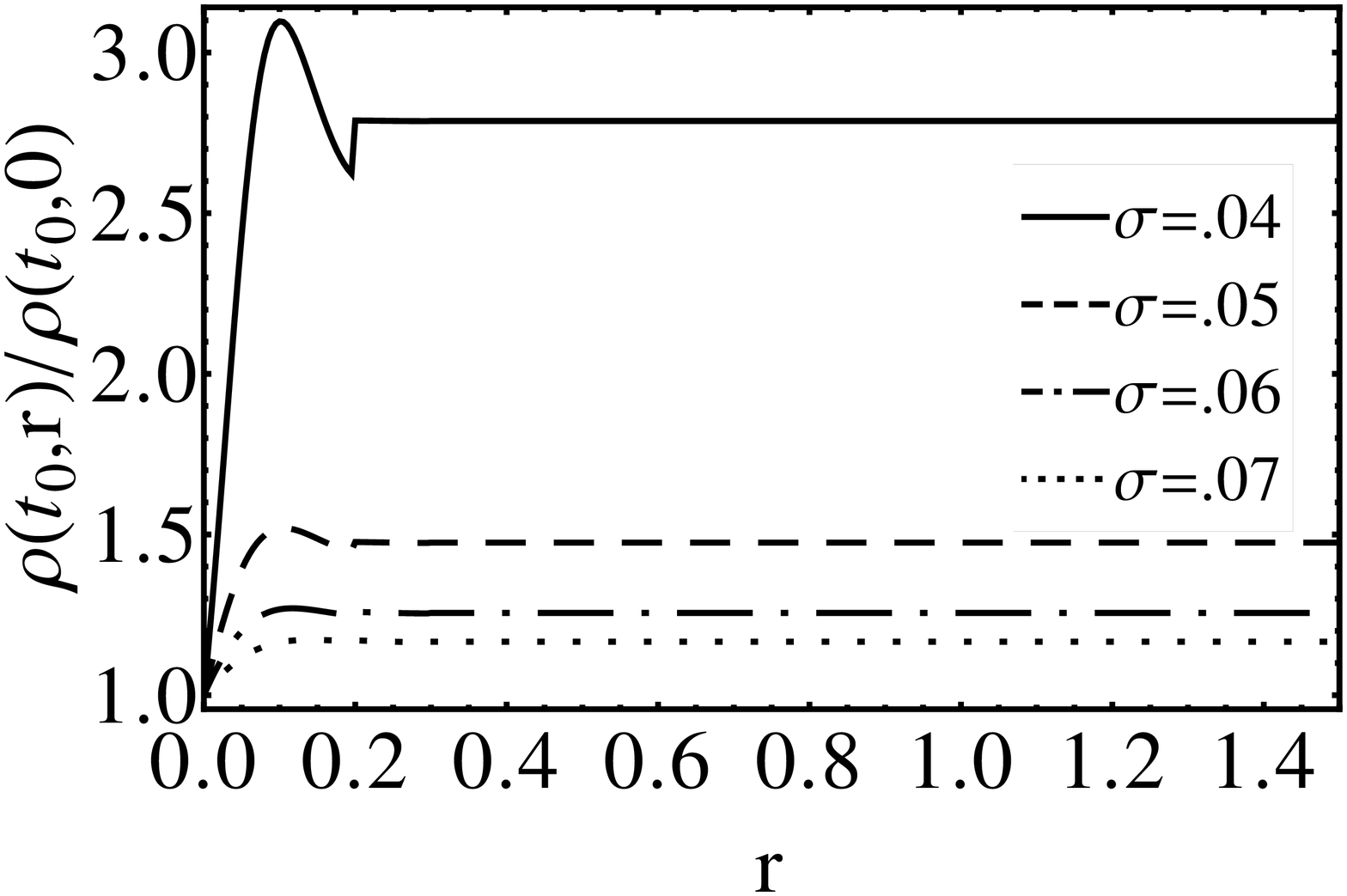}
\includegraphics[width= 0.5 \columnwidth]
{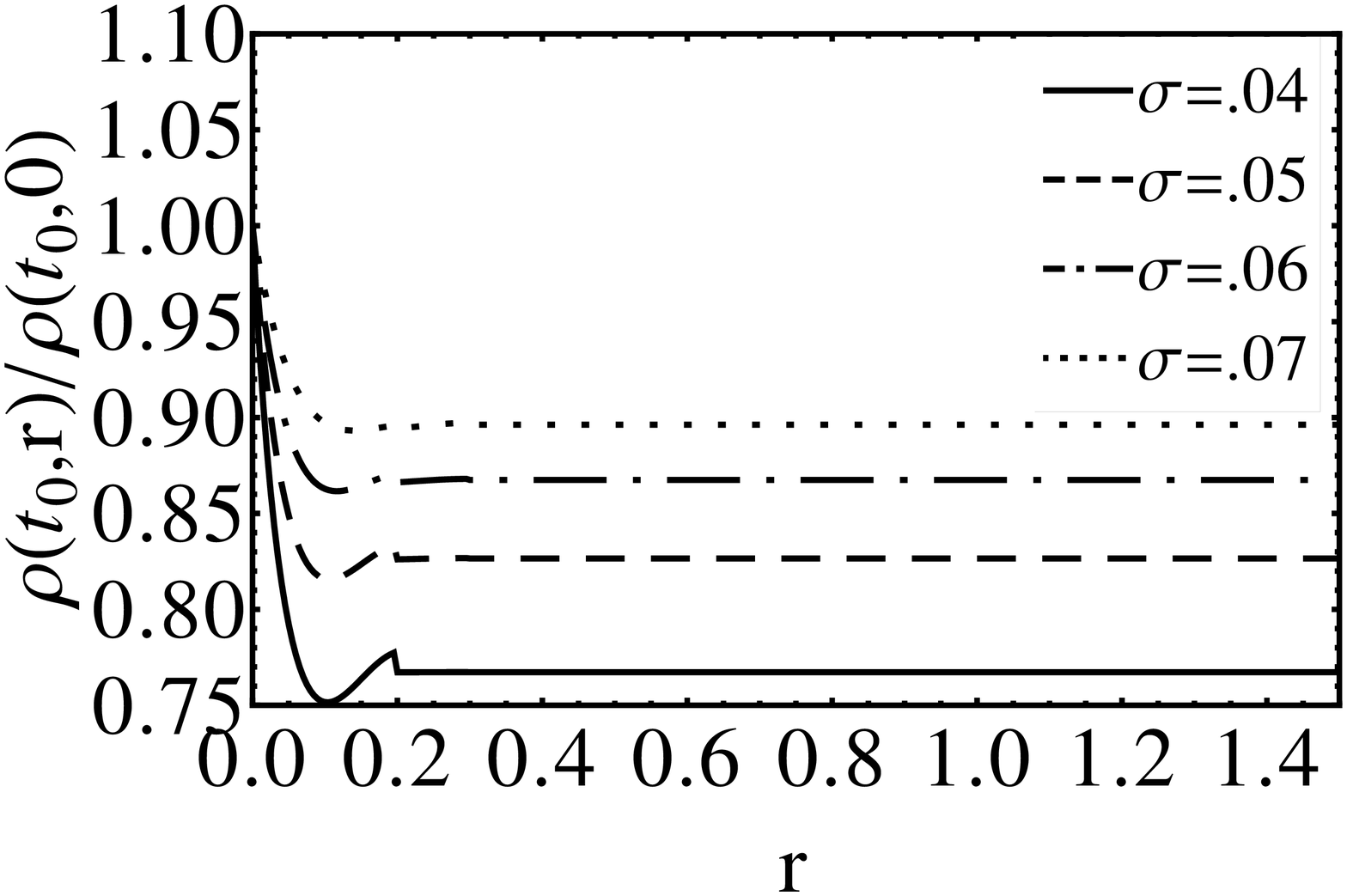}
\end{tabular}
\caption{The energy density ratio $\rho(t_0,r)/\rho(t_0,0)$ at the
time observation $t_0$  is plotted as function of the radial
coordinate for $A=-2\times(5\times   10^{-5})$ on the left and
$A=2\times(5\times   10^{-5})$ on the right. As can be seen
small values of $\sigma$ correspond to very large levels of
inhomogeneity, making them incompatible with observations. The
radial coordinate $r$ and  $\sigma$ are expressed in units of
$H_0^{-1}$.} \label{smalls2}
\end{figure}

\begin{figure}
\centering
\begin{tabular}{c}
\includegraphics[width=0.5 \columnwidth]
{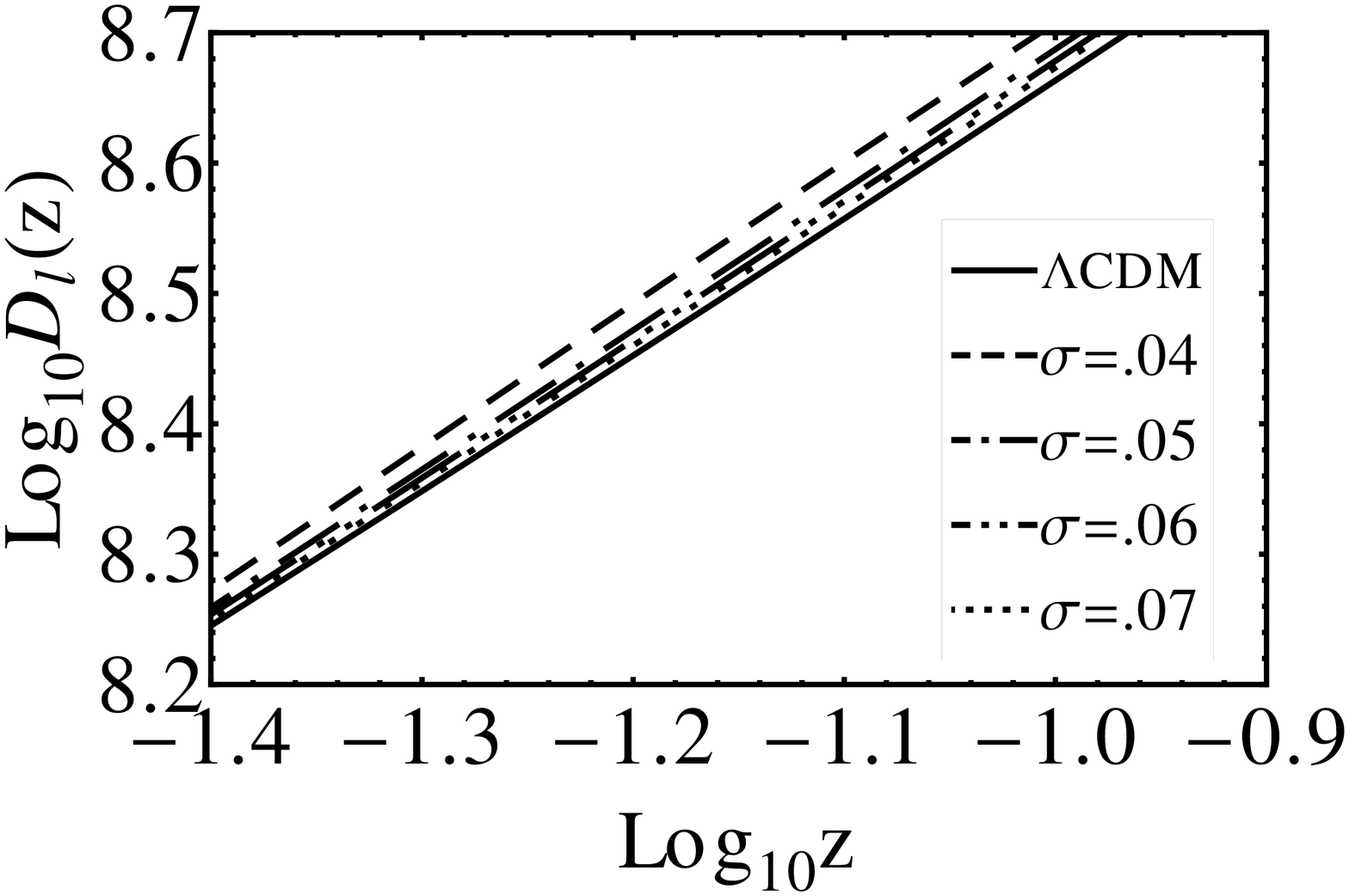}
\includegraphics[width=0.5 \columnwidth]
{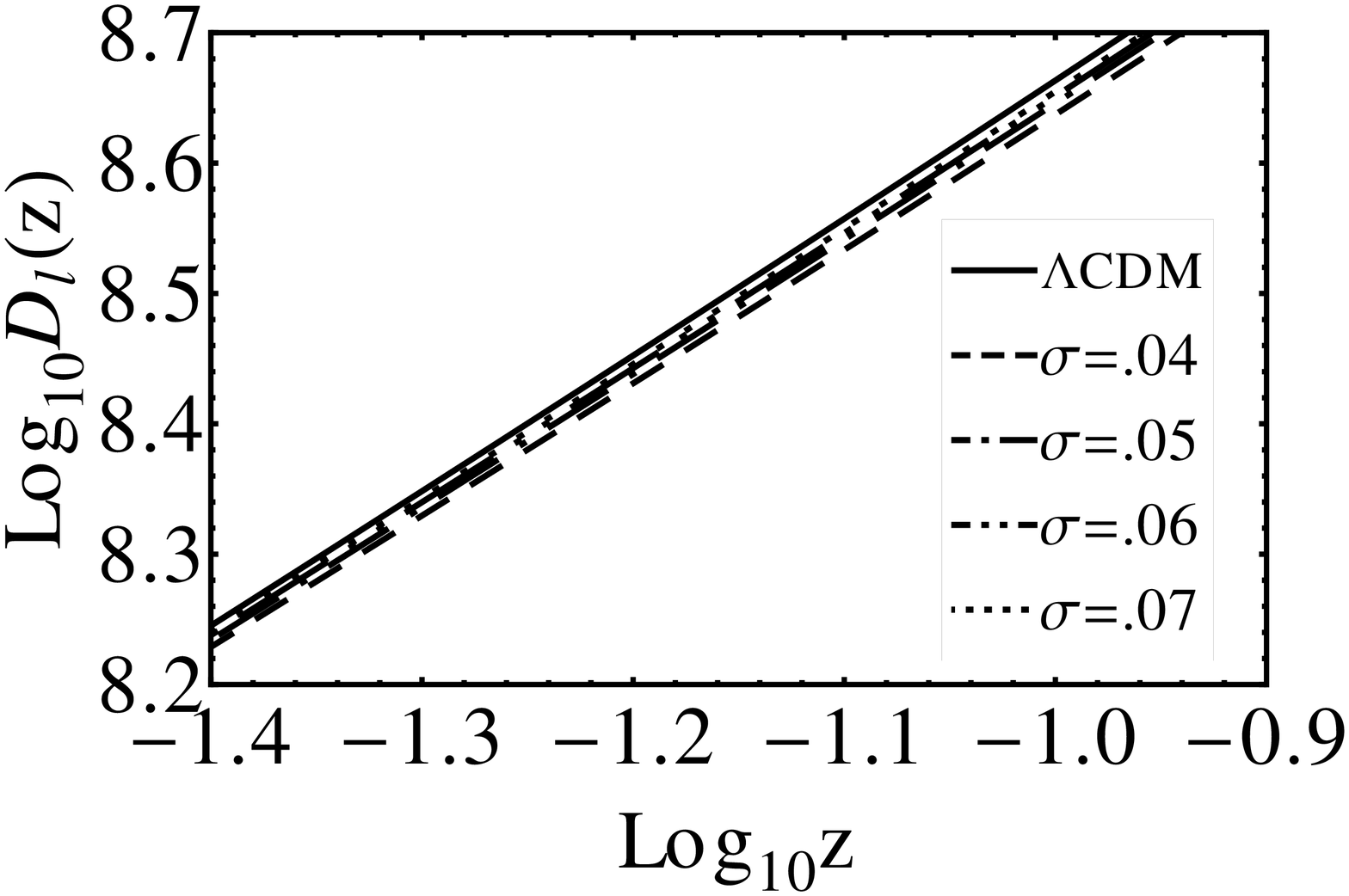}\\
\includegraphics[width=0.5 \columnwidth]
{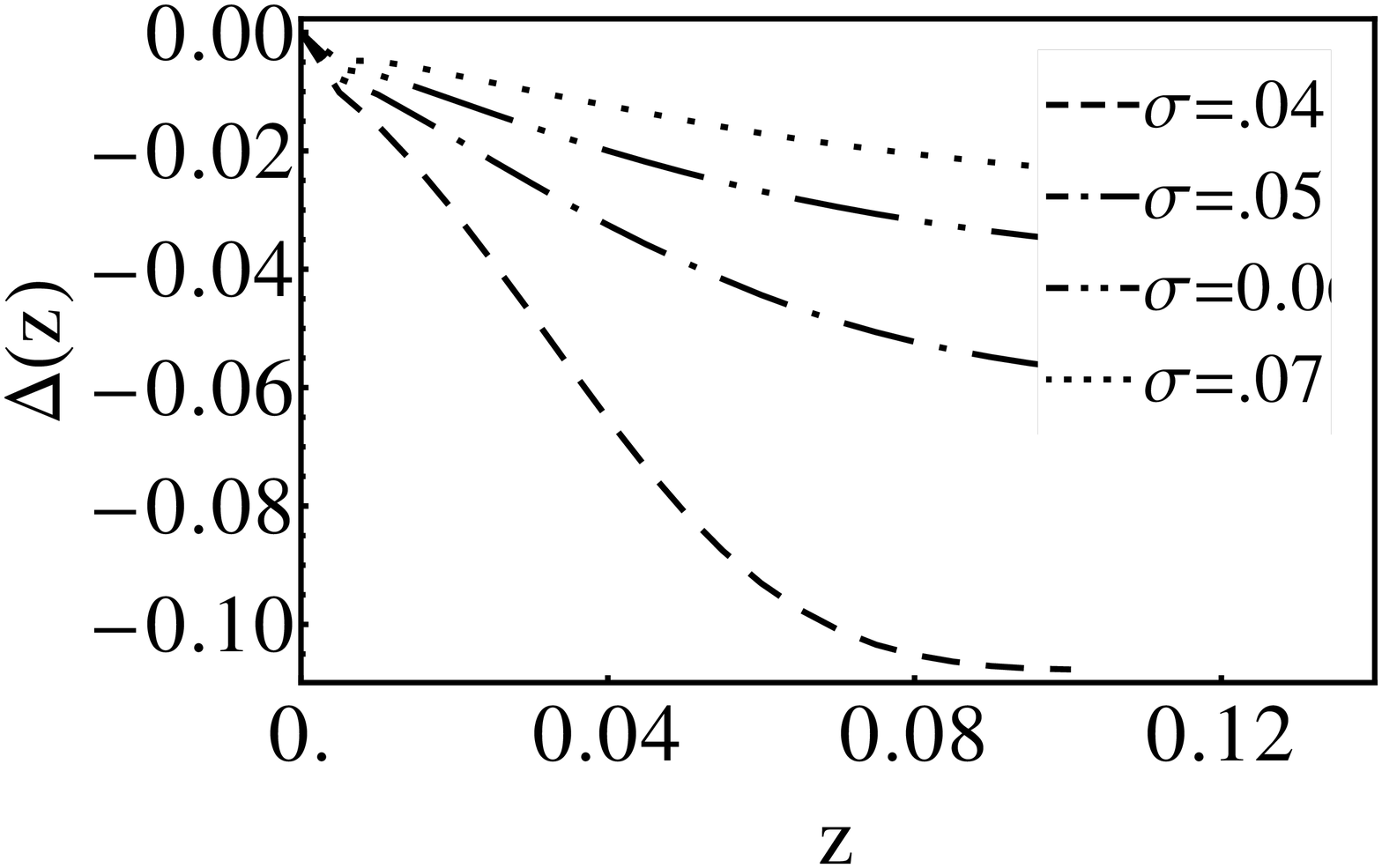}
\includegraphics[width=0.5 \columnwidth]
{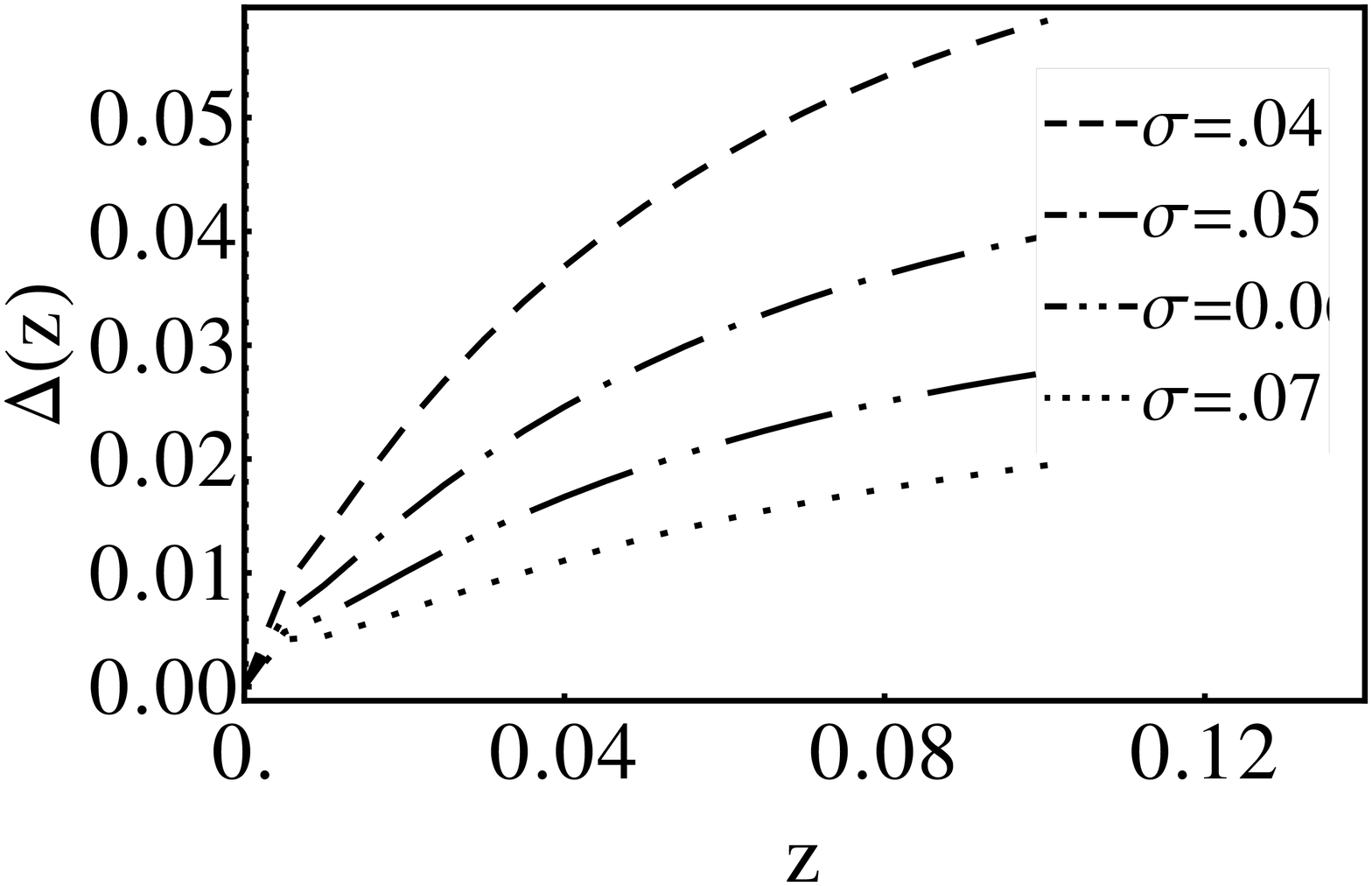} \\
\end{tabular}
\caption{On the top the luminosity distance is plotted for
different values of $\sigma$ in units of $H_0^{-1}$. On the bottom
the relative difference $\Delta(z)=(D_L^{\Lambda
CDM}(z)-D_L^{\Lambda LTB}(z))/D_L^{\Lambda CDM}(z)$ of the
luminosity distance between the LTB case and $\Lambda CDM$ is
plotted for different values of $\sigma$, in units of $H_0^{-1}$.
The left figures corresponds to $A=-2\times(5\times   10^{-5})$
and the right to  $A=2\times(5\times   10^{-5})$. These figures
correspond to small values of $\sigma$, showing how the
corresponding effective $H_0$ for $\sigma<0.04$, estimated as a
low redshift slope of the luminosity distance, is not compatible
with the observed value.  } \label{smalls}
\end{figure}
\begin{figure*}
\centering
\begin{tabular}{c}
\includegraphics[width=0.3 \columnwidth]{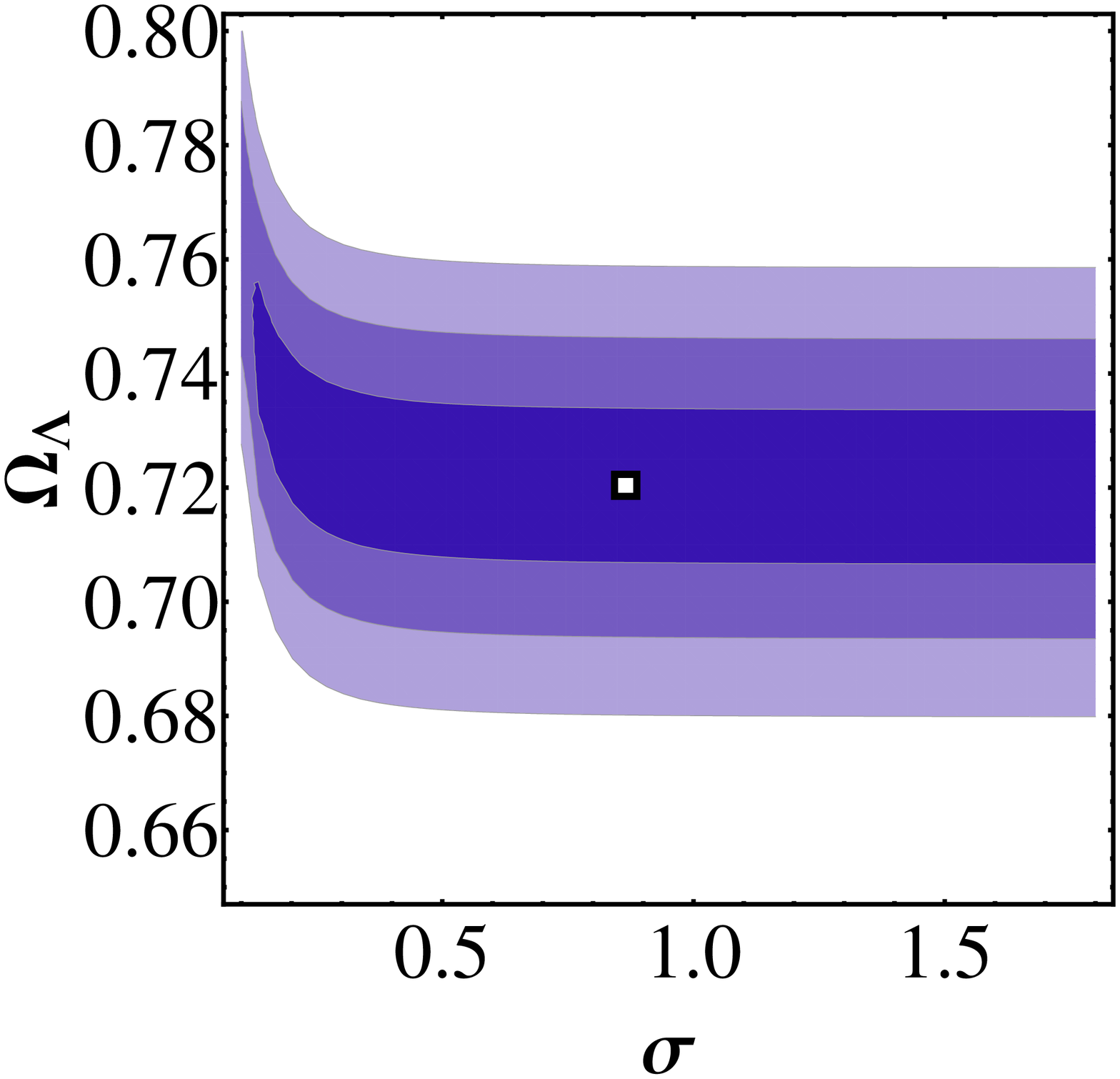}
\includegraphics[width=0.3 \columnwidth]{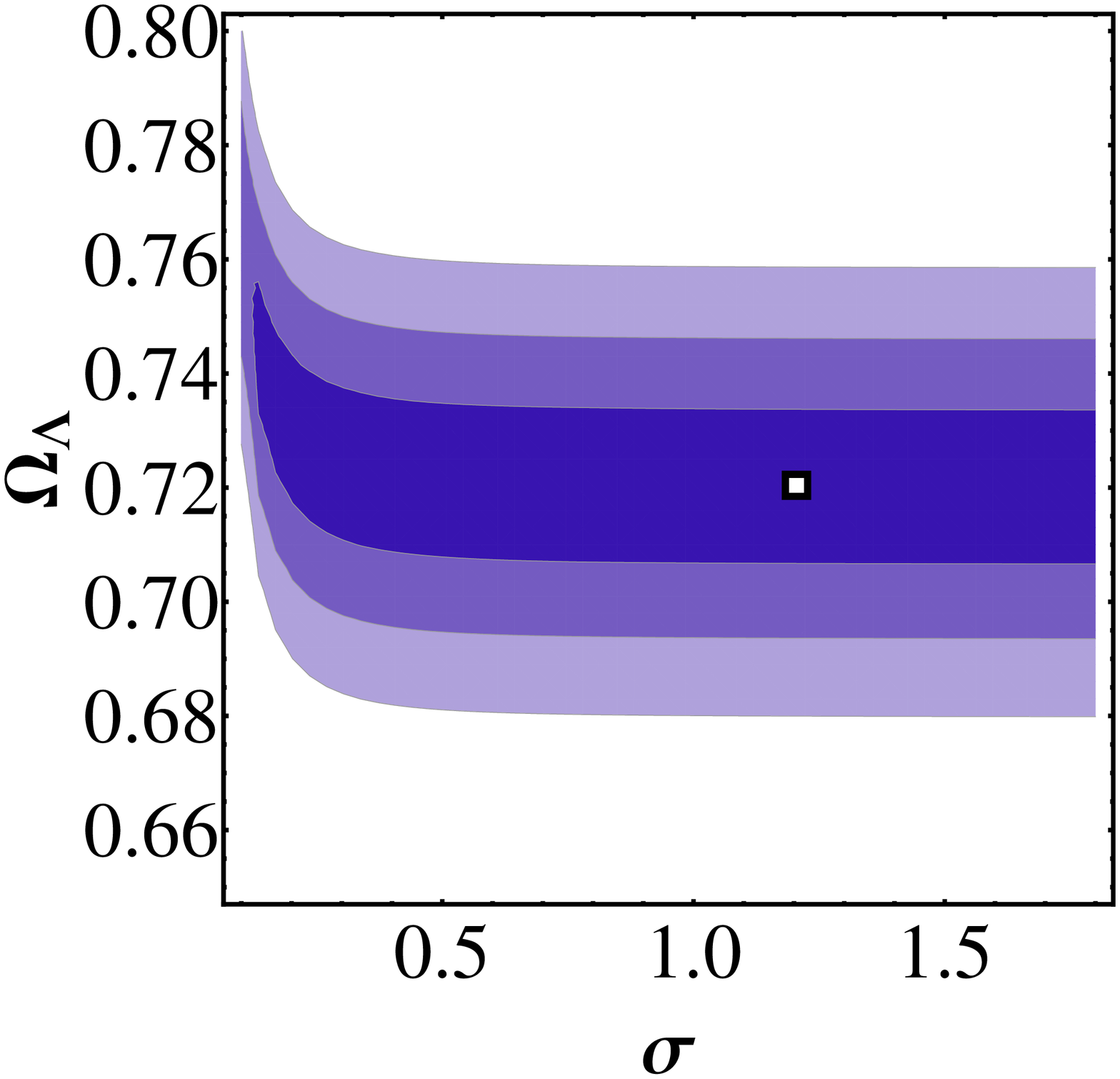}
\includegraphics[width=0.3 \columnwidth]{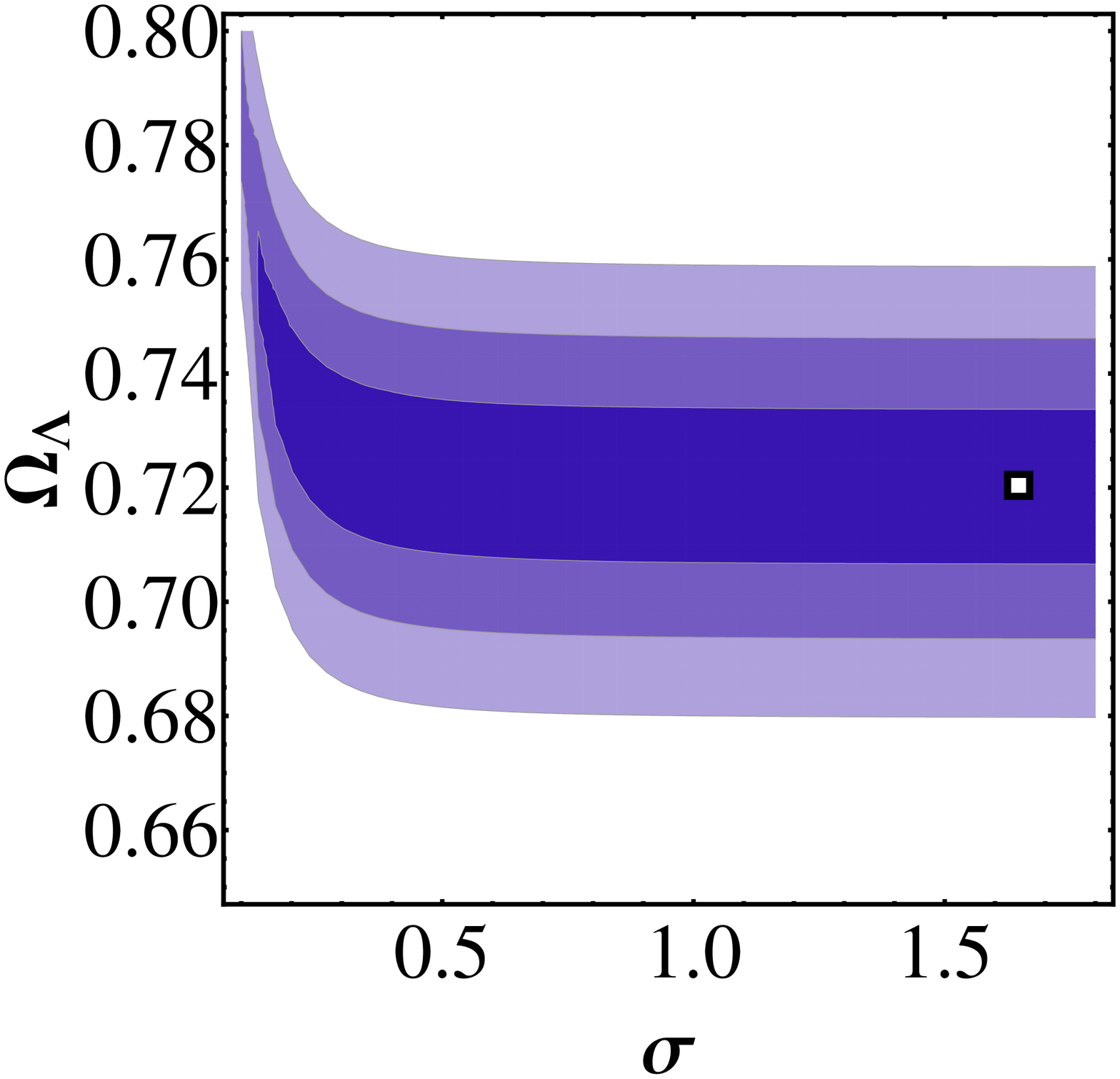}\\
\includegraphics[width=0.3 \columnwidth]{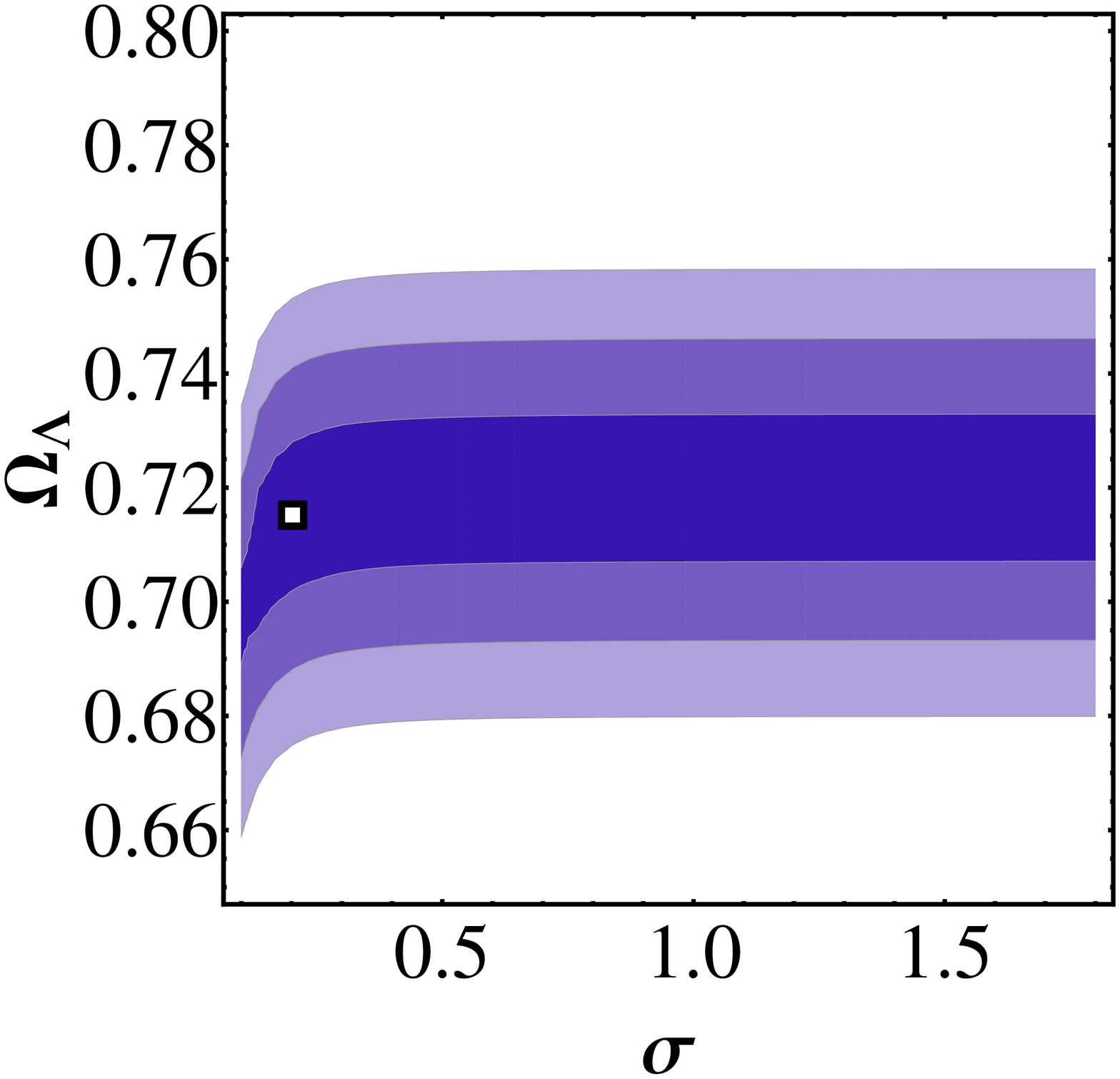}
\includegraphics[width=0.3 \columnwidth]{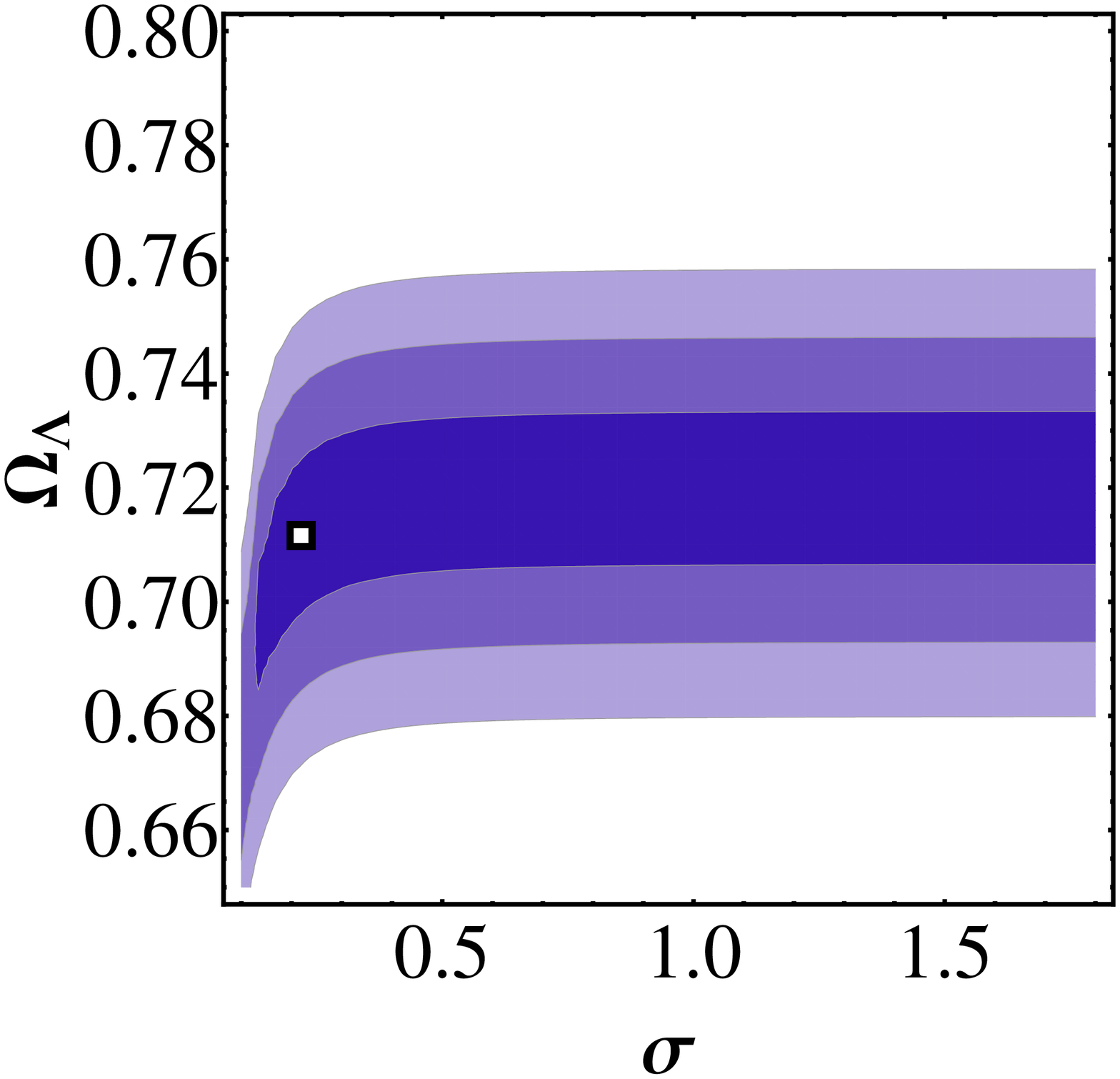}
\includegraphics[width=0.3 \columnwidth]{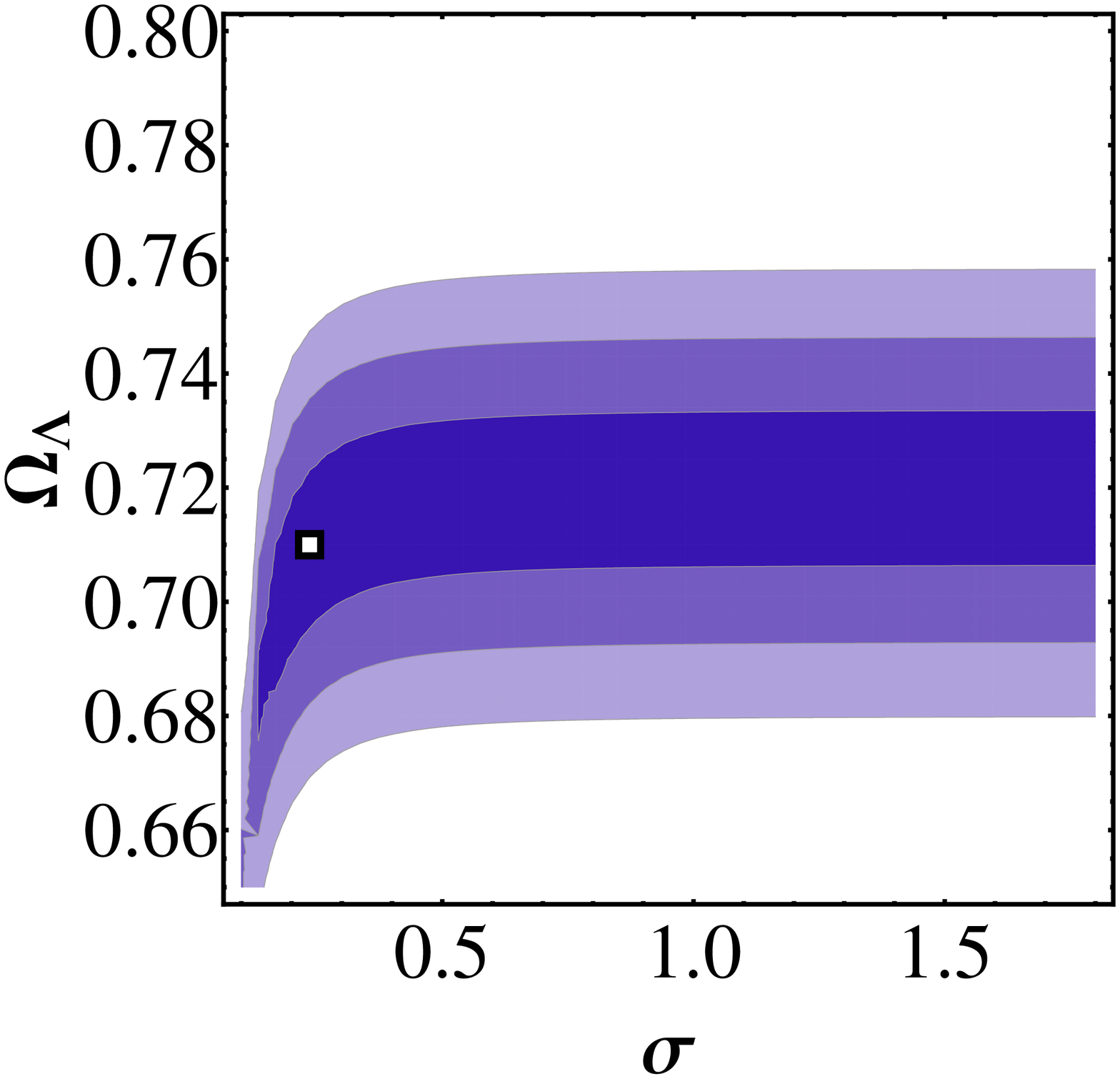}\\
\end{tabular}
\caption{The contour plots for the luminosity distance $\chi^2$
are shown for the parameters $\Omega_\Lambda$ and $\sigma$,
expressed in units of $H_0^{-1}$. For the top figures $A=1\times
5\times 10^{-5}$ , $A=2\times 5\times 10^{-5}$ and $A=3\times
5\times 10^{-5}$, from left to right respectively. For the bottom
figures $A=-1\times 5\times 10^{-5}$, $A=-2\times 5\times 10^{-5}$
and $A=-3\times 5\times 10^{-5}$, from left to right respectively.
} \label{contour}
\end{figure*}

\section{Effects on the estimation of the apparent value of the
cosmological constant from supernova Ia luminosity distance}
Our goal is to assess what could be the effect on the estimation
of the value of the cosmological constant due to a local
inhomogeneity seeded by a fluctuation of the primordial curvature
perturbation. In particular we will focus on the effects on the
supernovae Ia luminosity distance observations. In our analysis we
will use the Union $2.1$ compilation data set
\cite{Suzuki:2011hu}.

Given the assumption of the central location of the observer, we
need to solve the radial null geodesics  \cite{Celerier:1999hp}
\begin{eqnarray}
{dr\over dz}&=&{\sqrt{1+2E(r(z))}\over {(1+z)\dot
{R'}[T(r(z)),r(z)]}} \,,
\label{eq:34} \\
{dt\over dz}&=&-\,{R'[T(r(z),r(z))]\over {(1+z)\dot
{R'}[T(r(z)),r(z)]}} \,,
\label{eq:35} \\
\end{eqnarray}
and then substitute in the formula for the luminosity distance in
a LTB space \bea D_L(z)&=&(1+z)^2 R(t(z),r(z)) \,. \eea We will
model the primordial curvature perturbations  with a Gaussian
profile: \bea
 \zeta(r)=A e^{-(\frac{r}{\sigma})^2} \,,
\eea and fit data with the corresponding LTB solution given by
eq.(\ref{kz}). The initial conditions for the geodesic equations are obtained from the Einstein's equations at the center, and the age of the Universe used as initial condition for the time geodesics and reported in the last column of Table I  is obtained by integrating with respect to the scale factor from the big-bang till today. The parameter $\sigma$ is associated to the scale
of the inhomogeneity, while $A$ is related to its amplitude. From
observational constraints from the CMB anisotropy spectrum we know
that the standard deviation of the primordial curvature
perturbation $\zeta(r)$ should be about $5\times 10^{-5}$, so we
will consider peaks with $A$ equal to some  multiples of this
value. The relation between $k(r)$ and $\zeta(r)$ is shown in Fig.
\ref{fig-k-z}, for different values of $A$ and $\sigma$. The
present density contrast corresponding to peaks of the primordial
curvature perturbations with different values of $A$ and $\sigma$
is shown in Fig. \ref{rho}, while the effects on the luminosity
distance are shown in Fig. \ref{Dl}.As can be seen small values
of $\sigma$ are associated to larger $k(r)$, which also correspond
to greater observational effects. For this reason very small
values of $\sigma$ are observationally excluded, since they would
correspond to very high density contrasts today, and they would
also be incompatible with local measurements of the Hubble
parameter as shown in Fig. \ref{smalls}-\ref{smalls2}.
The results of the data fitting are shown in the contour plots for
the parameters ${\sigma,\Omega_{\Lambda}}$ in Fig. \ref{contour}.
As can be seen in Fig. \ref{rho} positive values of $A$
correspond to central overdensity, and negative values to
underdensities. The effects on the luminosity distance are shown
for both positive and negative curvature perturbations peaks in
Fig \ref{Dl}. For negative peaks we have an increase of the
luminosity distance with respect to a $\Lambda CDM$ with the same value
of $\Omega_{\Lambda}$, while for positive peaks the effect is the
opposite.

As it can be seen from the luminosity distance plots, values of
$\sigma$ lower than 0.1 introduce a large effect on the value of
$H_0$, making this area of the parameter space incompatible with
observational data of the Hubble parameter, but this does not
affect  our conclusions, since we find that the best fit
parameters correspond to $\sigma>0.2$.

\begin{table}
  \begin{tabular}{|c|c|c|c|c|}
   \hline
  $A/(5\times10^{-5})$ &  $\sigma$ & $\Omega_{\Lambda}$ & $\chi^2_{min}$ & $t_0$ \\
  \hline
  \hline
  3 & 1.64 & 0.7204  & 562.242 & 0.983023\\
  \hline
 2 & 1.212 &  0.7204 &562.242  & 0.982992\\
  \hline
  1 & 0.864 & 0.7204  & 562.242 & 0.982995 \\
  \hline
  0 &  &  0.72 & 562.242 & 0.982778\\
  \hline
  -1 &  0.209 & 0.7155  &  562.217  & 0.981357\\
  \hline
  -2 & 0.228 &  0.7124 & 562.202 & 0.980357\\
  \hline
  -3 & 0.232 & 0.709  & 562.190 & 0.979288\\
  \hline
  \end{tabular}
 \caption{The table shows the values of $\sigma$, expressed in units
 of $H_0^{-1}$, and $\Omega_{\Lambda}$ minimizing the $\chi^2$ for
 different values of the amplitude A, where the latter is expressed
 in integer multiples of the $5\times10^{-5}$, the value of the
 standard deviation of the primordial curvature perturbations implied
 by CMB observations. Positive values of A do not improve appreciably
 the value of $\chi^2$, neither affect greatly the best fit values for
 $\Omega_{\Lambda}$, while negative values improve the $\chi^2$ and
 also affect considerably the best fit value for $\Omega_{\Lambda}$.
 The row corresponding to $A=0$ gives the best fit values for a flat
 $\Lambda CDM$ model, and is reported as a reference for the magnitude
 of the effects of the local inhomogeneity. The last column corresponds to the age of the Universe in units of $H_0^{-1}$, showing how the effect on this is smaller than on the cosmological constant.}
 \label{TabI}
 \end{table}

From the table I we can see that the effects on the estimation of
the value of the cosmological constant can be of the order  of
$\approx 1.5\%$, which cannot be ignored, since they are of the
order of magnitude of other sources of systematic errors.

\section{Conclusions}
We have studied the effects of late time inhomogeneities seeded by
primordial curvature perturbations on the luminosity distance of
supernovae Ia, and consequently on the estimation of the apparent
value of a cosmological constant. Our analysis shows that these
effects cannot be ignored in the high precision cosmology era in
which we are entering and should be properly taken into account.
Fitting data under the a priori assumption of exact background
homogeneity could overestimate the value of the cosmological
constant by up to $1.5\%$. The same kind of conclusions can apply
to other cosmological parameters whose estimation can be affected
by local structure and suggests the importance of including into any cosmological
data analysis a realistic model of the local Universe which goes beyond the perturbative approach. In this paper we have considered the monopole contribution to these effects and we have related them to
primordial curvature perturbations of the type predicted by the
inflationary scenario.

\acknowledgments This work was supported in part by the
Grant-in-Aid for Scientific Research No.~21244033. A.S.\
acknowledges RESCEU hospitality as a visiting professor. He was
also partially supported by the grant RFBR 14-02-00894 and by the
Scientific Programme ``Astronomy'' of the Russian Academy of
Sciences.
AER was supported by Dedicacion exclusica and Sostenibilidad programs
at UDEA, and by the CODI project IN10219CE. SSN was supported by CODI
project IN615CE.


\end{document}